\documentclass[12pt]{article}

\usepackage{amsmath}
\usepackage{epsf}
\begin{document}
\sloppypar

 {\it Accepted for publication in Astronomy
Letters, v.  27, N 12, 2001.}
\vspace{2cm}
\bigskip
 
\large
\centerline{\bf Nodal and Periastron Precession of Inclined Orbits }
\centerline{\bf in the Field of a Rotating Black Hole .}
\vspace{15mm}
 \normalsize
\centerline{ Nail Sibgatullin$^{1,2}$ }
\vspace{2mm}
\noindent
$^1${\it Moscow State University, Vorob'evy gory, Moscow, 119899 Russia }\\
$^2$ {\it Max-Planck-Institut f\"ur Astrophysik, Karl-Schwarzschild-Str. 1,
85740 Garching bei \\ \noindent Munchen, Germany}\\

$^*$ {e-mail:sibgat@mech.math.msu.su}
 
\vspace{7mm}
\def\f{\frac}
\def\p{\partial}

\begin{abstract}
The inclination of low-eccentricity orbits is shown to significantly 
affect the orbital
parameters, in particular, the Keplerian, nodal precession, and 
periastron rotation
frequencies, which are interpreted in terms of observable quantities. 
For the nodal
precession and periastron rotation frequencies of low-eccentricity 
orbits in a Kerr
field, we derive a Taylor expansion in terms of the Kerr parameter at 
arbitrary orbital
inclinations to the black-hole spin axis and at arbitrary radial 
coordinates. The
particle radius, energy, and angular momentum in the marginally stable 
circular orbits
are calculated as functions of the Kerr parameter $j$ and parameter $s$ 
in the form of
Taylor expansions in terms of $j$ to within $O[j^6]$. By analyzing our 
numerical
results, we give compact approximation formulas for the nodal precession 
frequency of
the marginally stable circular orbits at various $s$ in the entire range of variation
of Kerr
parameter.

\end{abstract}

\section*{Introduction}

The X-ray quasi-periodic oscillations (QPOs) discovered in low-mass 
X-ray binaries
(LMXBs) commonly show a variety of modes. In particular, there are 
horizontal-branch
oscillations at low frequencies $\nu_{\mathrm{HBO}}$ 1--100~Hz and two 
peaks at
frequencies $\nu_1$ and $\nu_2$ ($\sim$1~kHz) in the power spectrum (van 
der Klis 2000).
In many Z-type sources, as well as in several atoll sources, the 
frequencies
$\nu_{\mathrm{HBO}}$ exhibit an almost quadratic dependence on $\nu_2$ 
(Stella and
Vietri 1998; Psaltis \emph{et al.} 1999). For several sources (e.g., for 
Sco~X-1,
4U~1608--52, 4U~1702--43, 4U~1735--44, XTE~J2123--058), the frequency 
difference
$\nu_2-\nu_1\equiv \Delta \nu$ decreases with increasing $\nu_2$. At the 
same time, for
many Z-type (GX~5-1,GX~17+2, Cyg~X-2, GX~340+0, GX~349+2) and atoll 
(4$\Gamma$~0614+09,
4$\Gamma$~1636-52, 4$\Gamma$~1705-44, Aq1~X-1) sources, the difference 
$\Delta \nu$ is
constant, which allows it to be associated with the stellar rotation 
frequency (van der
Klis 2000). The frequency $\nu_2$ is currently identified with the 
Keplerian rotation
frequency of clumps of matter near a compact object ($\nu_2\equiv 
\nu_{\mathrm{K}}$) in
almost all interpretations.

Cui \emph{et al.} (1998) pointed out that the nodal precession of 
circular
orbits\footnote{Lense--Thirring (1918) effect.} slightly inclined to the 
equatorial
plane of a black hole could be of importance in interpreting the stable 
QPOs: for several
black-hole candidates and microquasars, they found the predicted 
frequencies of nodal
precession to agree (if the Kerr parameter is chosen in the range 
$0.37-0.9$) with the
observed QPO frequencies.

Stella and Vietri (1998) justified the formula for the nodal precession 
frequency
$\nu_{\mathrm{nod}}$ of orbits slightly inclined to the equatorial plane 
and identified
its even harmonics with the frequency $\nu_{\mathrm{HBO}}$. The 
frequency $ \nu_1$ was
associated with the periastron rotation frequency $\nu_{\mathrm{per}}=
\nu_{\mathrm{K}}-\nu_r$ of a low-eccentricity orbit (a 
general-relativity effect). They
showed that the theoretical dependence $\nu_{r}(\nu_\phi)$ for an 
appropriately chosen
mass of the object in the range $1.8-2.2M_{\odot}$ was in good agreement 
with the
experimental points in the ($\nu_2 , \Delta \nu$) plane for various 
sources with evolving
frequencies. In the Kerr solution, the frequency $\nu_r$ becomes zero at 
the marginally
stable orbit. Therefore, when a radiating clump of matter passes from 
one Keplerian
orbit to another\footnote{Through turbulent viscosity and radiative 
deceleration.} by
increasing its rotation frequency, the frequency difference between the 
peaks
$\Delta\nu=\nu_r$ decreases and even approaches zero at the marginally 
stable orbit. This
tendency was observed for some of the sources (see above).

Morsink and Stella (1999), Stella and Vietri (1998), and Psaltis 
\emph{et al.} (1999)
showed the theoretical dependences 
$\nu_{\mathrm{nod}}(\nu_{\mathrm{K}})$ to agree with
the measured source frequencies in the $\nu_2,\nu_{\mathrm{HBO}}$ plane 
if
$\nu_{\mathrm{HBO}}$ was identified with an appropriate even harmonic of 
the nodal
frequency.

The proper rotation of the source (a neutron star or a black hole) 
remains a free
parameter, and it can be chosen by using the stable frequency observed 
in LMXBs in the
X-ray band during outbursts.

Another free parameter is the orbital inclination $s$ to the equatorial 
plane of a
neutron star or a black hole, which significantly affects the observed 
nodal precession,
periastron rotation, and Keplerian frequencies near the marginally 
stable orbit. Below,
we give an example based on our results. If the mass of a black hole or 
a neutron star is
$M$, then for the Keplerian frequency in the marginally stable orbit 
$\nu_{\phi}=1.2$
($2.2M_{\odot}/M$)~kHz, the frequency of nodal precession 
$\nu_{\mathrm{nod}}$ is,
respectively, 123 ($2.2M_{\odot}/M$)~Hz at $s=0$; $93.6$ 
($2.2M_{\odot}/M$)~Hz at $s=10^{\circ}$;
$65.3$ ($2.2M_{\odot}/M$)~Hz at $s=30^{\circ}$; $52.6$ 
($2.2M_{\odot}/M$)~Hz at
$s=45^{\circ}$; $41.18$ ($2.2M_{\odot}/M$)~Hz at $s=80^{\circ}$; and 
$41.11$
($2.2M_{\odot}/M$)~Hz at $s=90^{\circ}$. The nodal precession frequency 
changes by almost
a factor of~3 as the inclination changes from~$90^{\circ}$ to~0! Thus, by 
simultaneously
measuring three quantities, $M$, $\nu_{\mathrm{HBO}}$, and $\nu_2$, we 
can determine the
inclination of the marginally stable orbit to the spin axis!

In general, the tilt of an accretion disk to the equatorial plane of a 
compact object
can be finite. The Bardeen--Petterson (1975) hypothesis of the 
accretion-disk transition
into the equatorial plane breaks down  when the radiative forces 
that twist the
inner edge of the disk are taken into account. Pringle (1996) showed 
that a flat disk
was unstable to disturbances in the presence of a central radiation 
source.

Below, we derive formulas for the Keplerian, nodal precession, and 
periastron rotation
frequencies of low-eccentricity orbits at an arbitrary (finite) orbital 
inclination to
the equatorial plane in the form of Taylor expansions in terms of the 
Kerr parameter. In
contrast to previous studies, in which calculations were performed for 
selected values
of the constant~$Q$ [with an unclear physical meaning, as was noted by 
de~Felice (1980),
and coinciding with the square of the total angular momentum only in the 
weak-field
approximation $Q$), we consistently use the smallest latitudinal angle 
$\theta_-=s$
reached on a bounded trajectory\footnote{Below, $s$ is called the 
inclination angle
between the angular velocity vector of a compact object and the orbital 
surface for
short.} as a trajectory parameter. At $s=\pi/2$, the derived 
formulas~(51) transform into
Taylor expansions of the formulas by Okazaki \emph{et al.} (1987). An 
analysis of our
numerical results has yielded compact approximation formulas for the 
nodal precession
frequency of the marginally stable circular orbits in the entire range 
of Kerr parameters
for several finite orbital inclinations.

In contrast to rotating black holes, an intrinsic quadrupole component 
appears in the
external fields of rapidly rotating neutron stars (NSs). For NSs with a 
stiff equation
of state, this component can be several times larger than the Kerr 
quadrupole moment.
The effect of a non-Kerr NS field induced by rapid NS rotation on the 
parameters of the
marginally stable orbit and energy release in the equatorial boundary 
layer was analyzed
by Sibgatullin and Sunyaev (1998, 2000a, 2000b) using an exact solution 
for a rotating
configuration with a quadrupole moment (Manko \emph{et al.} 1994) 
[approximate
approaches with multipole expansions of the metric coefficients at large 
radii were
developed by Laarakkers and Poisson (1998) and Shibata and Sasaki 
(1998)]. Our results
refer to nonequatorial, nearly circular orbits in the fields of black 
holes and NSs with
a soft equation of state at a moderate rotation frequency. Markovic
(2000) considered finite-eccentricity orbits in a Kerr field and in the 
post-Newtonian
approximation for a field with a finite quadrupole moment. The 
precession of orbits
slightly inclined to the equatorial plane in the fields of rotating NSs 
was numerically
calculated by Morsink and Stella (1999) and Stella \emph{et al.} (1999).

\section*{Geodesics in the Kerr solution}

It is well known that the equations of geodesics in Riemannian spaces =
can be written in
Hamiltonian form. The corresponding Hamilton--Jacobi equation is
 \begin{gather}
 Q\equiv (g^{i,j} S,_i S,_j + 1)/2=
0,\qquad S,_i\equiv \frac{\partial S}{\partial x_i},\\
i, j,...= 0, 1, 2, 3.\nonumber
\end{gather}
The generalized momenta $p_i = S,_i$ are related to the 4-velocity 
components by the
Hamilton equations
\begin{gather}
 \frac{dx^j}{d\tau}=\frac{\partial Q}{\partial p_j}= g^{jk}p_k, \;
 \frac{dp_j}{d\tau}= -\frac{\partial
Q}{\partial x^j} = -\frac{\partial g^{kl}}{\partial x^j} p_k p_l.
\end{gather}
For the Kerr solution in Boyer--Lindquist coordinates, the contravariant 
metric
components are
\begin{gather}
g^{00}= -\frac{(r^2+a^2)^2-\Delta a^2\sin^2{\theta}}{\rho^2
c^2\Delta},\\
g^{\phi\phi}=\frac{(1-2rGM/c^2\rho^2)}{\Delta\sin{\theta}},\nonumber
\\
g^{0\phi}=\frac{-2ar}{\rho^2\Delta},\quad
g^{rr}=\frac{\Delta}{\rho^2},\quad
g^{\theta\theta}=\frac{1}{\rho^2}.\nonumber
\\
\Delta\equiv r^2+a^2 -2rGM/c^2;\quad \rho^2\equiv
r^2+a^2\cos^2{\theta}.\nonumber
\end{gather}
The constant~$a$ is related to the angular momentum of a rotating black 
hole described
by the Kerr solution by the equality $J= M a c = G M^2 j/c$, where 
$j$ is the
dimensionless Kerr parameter. Because of axial symmetry and 
stationarity, the
Hamilton--Jacobi equation~(1) for the Kerr metric has two cyclic 
coordinates, $t$ and
$\phi$; therefore, Eq.~(1) with the first integrals $p_t =
-E=\mbox{const}$ and
$p_{\phi}=L=\mbox{const}$ for the Kerr metric can be written as
\begin{gather}
(1-E^2)((r^2+a^2)^2-\Delta a^2\sin^2{\theta})+\\
 4 a r L E GM/c^3 +
(\rho^2-2GMr/c^2)\times\nonumber\\
 L^2/(c\sin{\theta})^2 \Delta^2 S,_r^2+\Delta S,_{\theta}^2 =
2r(r^2+a^2)GM/c^2.\nonumber
\end{gather}
The existence of the complete integral of the Hamilton--Jacobi equation 
established by
Carter~(1968) is less obvious:
\begin{gather*}
S=-Et c^2+L\phi+\int\sqrt{\Theta(\theta)}d\theta +
\int\sqrt{R(r)}\frac{dr}{\Delta},
\end{gather*}
\begin{gather}
\Theta(\theta)\equiv Q-\cos^2{\theta}(a^2 c^2 (1-E^2) + L^2/(\sin{\theta})^2),\\
R(r)\equiv (c E(r^2+a^2)-L a)^2-\\
\Delta (r^2+(L-a E c )^2 +Q.\nonumber
\end{gather}
Instead of the constant $Q$, we introduce a constant $s < \pi/2$, which 
has the meaning
of the minimum angle $\theta$ (turning point) reached on the bounded 
trajectory in
question [in the notation of Wilkins (1972), the constant $s\equiv 
\theta_-$; see also
Shakura (1987)]. According to (5), the constant $Q$ can be expressed in 
terms of $s$:
\begin{gather}
Q=\cos^2{s}(a^2 c^2(1-E^2) + L^2/(\sin{s})^2).
\end{gather}
In that case, the angle~$\theta$ on a bounded trajectory varies over the 
range $ s <
\theta < \pi - s $.

For stable trajectories on the $r=\mathrm{const}$ surfaces, the 
following conditions
must be satisfied:
\begin{gather}
R(r)=0,\quad dR(r)/dr=0
\end{gather}
(Bardeen \emph{et al.} 1972; Wilkins 1972).

\section*{The newtonian analog of the Kerr solution}

\subsection*{The newtonian analog of the supercritical Kerr solution and 
its potentials}

In the Newtonian limit, $ E^2\approx 1+2 H/c^2$ and $\Delta \approx 
r^2+a^2$.
Substituting these expressions in Eq.~(4) yields an approximate 
Hamilton--Jacobi
equation:
\begin{gather}
2H=\frac{
L^2}{(r^2+a^2)\sin^2{\theta}}+\frac{r^2+a^2}{r^2+a^2\cos^2{\theta}}
S,_r^2+\\
\frac{1}{r^2+a^2\cos^2{\theta}} S,_{\theta}^2+ \frac{
4arLGM/c}{(r^2+a^2)(r^2+a^2\cos^2{\theta})}
 -\nonumber\\
 \frac{2 GM
r}{r^2+a^2\cos^2{\theta}}+\frac{4(
GMra\sin{\theta})^2}{(r^2+a^2\cos^2{\theta})c^2(r^2 + a^2)}.\nonumber
\end{gather}

Note that Eq.~(8) describes the trajectories of test particles in the 
field of a flat
disk with radius~$a$ in special curvilinear coordinates $r$, $\theta$, 
$\phi$, which are
related to Cartesian coordinates $x, y, z$ by
\begin{gather}
r = \frac12(\sqrt{x^2+y^2 +(z-ia)^2}+{}\\
{}+\sqrt{x^2+y^2 +(z+ia)^2}),\notag
\end{gather}
\begin{gather}
 a\sin{\theta}=\frac{1}{2i}(\sqrt{x^2+y^2 +(z+ia)^2}-\\
 \sqrt{x^2+y^2+(z-ia)^2}),\quad \phi=\arctan{y/x}.\nonumber
\end{gather}
When passing from Cartesian coordinates $x,y,z$ to orthogonal 
curvilinear coordinates
$r,\theta,\phi$ using formulas~(10) and~(11), the nonzero metric tensor 
components are
expressed in terms of the curvilinear coordinates as
\begin{gather}
g_{rr}=\frac{r^2+a^2\cos^2{\theta}}{r^2+a^2},\quad
g_{\theta\theta}=r^2+a^2\cos^2{\theta},\\
g_{\phi\phi}=(r^2+a^2)\sin^2{\theta}.\nonumber
\end{gather}
Hamiltonian~(9) contains the contravariant metric tensor 
components~(12), and it can be
written in Cartesian coordinates as
\begin{gather*}
H=\frac{({p}+{\mathbf{A}}/c)^2}{2} -\Phi.
\end{gather*}
The corresponding Hamilton equations are
\begin{gather}
\frac{d p_k}{dt} =  \Phi,_k - ({p}+{\mathbf{A}}/c)
\frac{\partial{\mathbf{A}}/c}{\partial
x^k},\quad\frac{d{x^k}}{dt}=p_k + \mathbf{A}_k/c.
\end{gather}
Eqs.~(13) can be rewritten as the equations of motion in a 
gravitomagnetic field in the
quasi-Newtonian approximation:
\begin{gather}
\frac{d {v}}{d t}=\nabla \Phi +\mbox{curl}{ \mathbf{A}}\times { v}/c,
\end{gather}
where the Newtonian potential $\Phi$ and the gravitomagnetic vector 
potential
${\mathbf{A}}$ in a vacuum satisfy the Laplace equation. The equations 
of relative
motion in a rotating coordinate system can be derived from Eqs.~(14) if
$\boldsymbol{\Omega}\times{\mathbf{R}}$, ${\mathbf{R}}=(x,y,z)$, is 
substituted for
${\mathbf{A}}$. In that case, 
$\mbox{curl}{\mathbf{A}/c}=-2{\boldsymbol{\Omega}}$ and
the second term on the right-hand side of~(13) represents the Coriolis 
force.

For a flat Kerr disk, the Newtonian potential $\Phi $ is (Israel 1970; 
Zaripov \emph{et
al.} 1995)
\begin{gather}
\Phi= \frac{GM}{2\sqrt{x^2+y^2 +(z+ia)^2}}+ {}\\
\frac{GM}{2\sqrt{x^2+y^2
+(z-ia)^2}}. \notag
\end{gather}
The vector~${\mathbf{A}}$ consists of the components
$-g_{0\alpha}/g_{00},\quad\alpha=1,2,3$ (Landau and Lifshitz 1980). In 
the Newtonian
limit in a vacuum, the vector of the gravitomagnetic field~$\Psi$ can be 
introduced
instead of the vector potential~${\mathbf{A}}$:
\begin{gather}
\mbox{curl}{\mathbf{A}}=2 \nabla \Psi.
\end{gather}
The scalar~$\Psi$ in an axisymmetric case matches the imaginary part of 
the Ernst
complex potential. For the special case of a Kerr disk, the 
scalar~$\Psi$ is (Zaripov
\emph{et al.} 1995):
\begin{gather}
\Psi= \frac{GM}{2i\sqrt{x^2+y^2 +(z-ia)^2}} -{} \\
\frac{GM}{2i\sqrt{x^2+y^2 +(z+ia)^2}}. \notag
\end{gather}
Clearly, expressions~(15) and~(17) for $\Phi$ and $\Psi$ satisfy the 
Laplace equation.

As follows from definition~(10), the $r=\mbox{const}$ surface is 
neither a sphere nor an
ellipsoid.

\subsection*{The precession of circular keplerian Orbits in the field of 
a gravitating rotating mass}

For the Newtonian analog of a supercritical Kerr disk, \mbox{$a> 
GM/c^2$}. However,
since \mbox{$a<GM/c^2$} for black holes, in this subsection, we discard 
terms of the
order of $a^2$ in expressions~(15) and~(17) for $\Phi$ and $\Psi$ lest 
the order of
accuracy of the model be exceeded. In that case, $\Phi\approx GM/r,\quad 
\Psi\approx - a
z GM/r^3$.

Let us introduce a coordinate system with the origin at the gravitating 
mass with the
$z$~axis directed along the spin axis. The equations of motion for free 
particles in
this coordinate system are
\begin{gather}
 \frac{d u}{d t}= - \frac{GM x}{r^3} +
 \left(6a(yw-zv)\frac{z}{r^5}+2v\frac{a}{r^3}\right)\frac{GM}{c},
\\
 \frac{d v}{d t}= - \frac{GM y}{r^3}+\left(6a(z u- x w)\frac{z}{r^5}
 - 2u\frac{a}{r^3}\right)\frac{GM}{c},
\\
 \frac{d w}{d t}= - \frac{GM z}{r^3}+\left( 6a(x v- y =
u)\frac{z}{r^5}\right)\frac{GM}{c}.
\end{gather}

In our approximation, the complete integral of the Hamilton--Jacobi 
equation is [cf.
formulas (5)--(6)]
\begin{gather*}
S=-Ht+L\phi+\int\sqrt{\Theta(\theta)}d\theta + \int\sqrt{2 R(r)}dr,
\end{gather*}
\begin{gather}
\Theta(\theta)\equiv L^2\cot^2{s}-L^2\cot^2{\theta}.
\\
R(r)\equiv H+\frac{GM}{r}-\frac{L^2}{2r^2\sin^2{s}}-2\frac{LaGM}{r^3
c}.
\end{gather}
In formulas~(21) and~(22), we chose the turning point for the angle 
$\theta$ as a
constant: according to~(20), $\theta$ varies over the range $s<\theta< 
\pi-s$, with
\mbox{$ 0 < s < \pi/2$}.

For stable circular trajectories, the equalities $R(r)=0$ and $\quad 
dR(r)/dr=0$ must be
satisfied. Substituting expression~(21) for a fixed radius~$r$ in these 
equations yields
the corresponding energy~$H$ and angular momentum~$L$:
\begin{gather}
H\approx-\frac{GM}{2r}-a\sqrt{(GMr)^3}\frac{\sin{s}}{c r^4},\\
L\approx \sin{s}\sqrt{GMR}- 3aGM\frac{\sin^2{s}}{cr}.\nonumber
\end{gather}
Consider the precession of circular orbits. From the Hamilton equations
$\dot{\theta}=\partial H/\partial p_{\theta}$ and =
$\dot{\phi}=\partial H/\partial L$, we
have
\begin{gather}
\frac{d\theta}{d t}=\frac{L}{r^2}\sqrt{\cot^2{s}-\cot^2{\theta}},
\\
\frac{d\phi}{d t}= \frac{L}{r^2\sin^2{\theta}} +\frac{2aGM}{r^3c}.
\end{gather}
Integrating Eq.~(24) yields
\begin{gather}
\phi-\phi_0=\int\frac{L dt}{r^2\sin^2{\theta}}+\frac{2aGMt}{r^3
c}=\\
\int\frac{d\theta}{\sin^2{\theta}\sqrt{\mathrm{c}^2{s}-\mathrm{c}^2{
\theta}}}+\frac{2aGMt}{r^3c}
=\nonumber\\
\arcsin{\left(\frac{\cot{\theta}}{\cot{s}}\right)}+\frac{2aGMt}{r^3c
}.\nonumber
\end{gather}
When changing the variable in formula~(26), we used Eq.~(24).

Consider the case where all particle at $t=0$ were in the
$\sin{\phi}\sin{\theta}=\cos{\theta}\cot{s}$ plane or, introducing 
Cartesian
coordinates, in the $y=z\cot{s}$ plane. According to equality~(26), 
the particles will
lie on the $y\cos{(2aGMt/r^3c)}-x\sin{(2aGMt/r^3c)}\,=\,z\cot{s}$ 
surface at time~$t$.
The closer the particles to the rotating gravitating center, the faster 
their precession.
For a given radius~$r$, the particle motion in an accretion disk can be 
represented as
the rotation of a circumference inclined to the $z$ spin axis at 
angle~$s$ with angular
velocity~$2aGM/r^3c$. In a coordinate system corotating with the 
circumference, the
particle rotates with a Keplerian velocity. Thus, if viscous friction 
between adjacent
orbits is disregarded, a tilted accretion disk cannot be in a steady 
state.

\section*{The energy and angular momentum of particles moving along
geodesics orbits \\ \protect in the Kerr solution on the $r=\mbox{const}$ surfaces}

In this section, for simplicity, we choose a system of units in which 
$c= G=M=1$. To
find stable orbits on the $r=\mbox{const}$ surfaces, we must solve the 
algebraic
equations~(8). The corresponding solutions in the $s=\pi/2$ equatorial 
plane were found
by Ruffini and Wheeler (1970) for $j=1$ and by Bardeen \emph{et al.} 
(1972) for
arbitrary $j$. For $s=0$ and arbitrary $j$, the particle angular 
momentum~$L$ is zero.
The corresponding expression for the energy was derived by Lightman 
\emph{et al.} (1975).
Inclined orbits at $j=1$ were investigated by Wilkins (1972). In the case
of charged rotating black holes the selected tilted bound geodesics were studied 
by Johnston and Ruffini in extreme case $j=1$. In the case
of charged rotating black holes the selected tilted bound geodesics were studied 
by Johnston and Ruffini in extreme case $j=1$. Finally, 
Shakura (1987)
derived formulas for the energy and angular momentum at arbitrary $j$ 
and $s$ by a
complex, indirect method. Below, we show that derivation expressions for 
$E$ and $L$ from
Eqs.~(8) can be reduced  to solving a quadratic equation and factorizing the 
numerator and
denominator in the resulting fraction (to be subsequently canceled by a 
common factor).

Let us introduce new unknowns instead of $E$ and $L$:
\begin{gather}
x_c\equiv \frac{E}{L},\quad y_c= \frac{E^2-1}{L^2}.
\end{gather}
Eqs.~(8) can then be written as
\begin{gather}
y_c a_1+x_c a_2 +(x_c^2-y_c)(a_1+a_3)+a_0=0,
\end{gather}
\begin{gather}
y_c b_1+x_c b_2 +(x_c^2-y_c)(b_1+b_3)+b_0=0, \\
 b_i\equiv
\frac{\partial a_i}{\partial r}|_{j,s},\quad i=0,1,2,3.\nonumber
\end{gather}
Here,
\begin{gather*}
a_0=-(r^2-2r +\Delta \cos^2{s}),\\
a_1=r^4+2j^2+j^2r+j^2\Delta\cos^2{s},
\end{gather*}
\begin{gather}
 a_2=-4jr,\quad
a_3=-p\Delta,\; p\equiv r^2+j^2\cos^2{s}.
\end{gather}
Eliminating $y$ from Eqs.~(28) and~(29) yields the quadratic equation
\begin{gather*}
x_c^2 (a_1 b_3-a_3 b_1)+x_c(a_2 b_3-a_3 b_2)+{}\\
{}+a_0 b_3-a_3 b_0 = 0,
\end{gather*}
whose solution is given by
\begin{gather}
x_c=\frac{E}{L}=\frac{A+q B}{D\sin{s}},\quad
q\equiv\sqrt{r-j^2\cos^2{s}/r}.
\end{gather}
In formula~(31), we use the notation
\begin{gather}
A\equiv j\sin{s}(3r^4-4r^3 + j^2 r^2\cos^2{s}(j^2r^2-r^4)),\nonumber\\
B\equiv rp\Delta,\\
D=-a_1b_3+a_3b_1=(r^2-j^2\cos^2{s})\times\\
(r^2+j^2)^2-4j^2r^3\sin^2{s}.\nonumber
\end{gather}
The numerator and denominator in~(31) can be factorized as
\begin{gather}
A+q B=r((r^2+j^2)q+2jr\sin{s})\times\\
(j\sin{s}q+r^2+j^2\cos^2{s}-2r),\nonumber
\\
D=r((r^2+j^2)q+2jr\sin{s})((r^2+j^2)q-2jr\sin{s}).
\end{gather}
Therefore, cancelling a common factor in the numerator and denominator 
in~(31), we obtain
\begin{gather}
x_c=\frac{E}{L}=\frac{jq\sin{s} + r^2+j^2\cos^2{s}-2r}{(r^2+j^2)
q-2jr\sin{s})\sin{s}}.
\end{gather}
Multiplying Eq.~(28) by $b_1$ and Eq.~(29) by $a_1$  and 
subtracting~(29) from~(28)
yields
\begin{gather}
x_c^2-y_c=\frac{1}{L^2}=\frac{x(a_1 b_2-a_2 b_1)+a_1 b_0-a_0
b_1}{D}\\=\frac{A_1+q B_1}{D\sin{s}},\nonumber
\end{gather}
\begin{gather*}
A_1\equiv 2jr p(2r^4-3r^3+j^2r^2+\cos^2{s} j^2(r-j^2)),\\
 B_1\equiv
p^2(r^3-3r^2+j^2+j^2r).
\end{gather*}
The expression $A_1+q B_1$ can also be factorized as
\begin{gather}
A_1+q B_1= p ((r^2+j^2)q+2jr\sin{s})\times\\
(2 r j q \sin{s} + r^3-3 r^2 + j^2\cos^2{s}(r+1)).\nonumber
\end{gather}
Hence,
\begin{gather}
L=\sqrt{\frac{D\sin{s}}{A_1 + q B_1}}={}\\
\frac{((r^2+j^2)q - 2 j
r\sin{s})\sin{s}}{\sqrt{ p ( p - 3 r + j^2\cos^2{s}/r + 2 j q  r
\sin{s})}}.\nonumber
\end{gather}
Using formula~(36) for $E/L$, we obtain with~(39)
\begin{gather}
E = \frac{p-2 r +j q \sin{s} }{\sqrt{p ( p - 3 r + j^2\cos^2{s}/r +
2 j q  r \sin{s} )}}.
\end{gather}
Formulas~(39) and~(40) are the sought-for expressions for the particle 
energy and
angular momentum in inclined circular orbits.

We determine the coordinate radius of the marginally stable orbit, 
$r_*$, from the
condition of $E$ given by~(40) being at a minimum. Representing~$r_*$ as 
a Taylor
expansion in terms of $j$, we obtain $r_*$ as a function of $j, s$:
\begin{gather}
r_*\approx 6-4\sqrt{\frac23} j \sin{s}+..\approx 6-3.266 j \sin{s}
-\\
j^2 (0.5 + 0.1111 \cos{2s}) +j^3 (-0.2532 \sin{s} 
-\nonumber\\
 0.0567 \sin{3s}) + j^4 (-0.1196+ 0.0206 
\cos{2s}+\nonumber\\
 0.0162 \cos{4s})+j^5 (-0.11 \sin{s} - 0.02 \sin{3s}
+\nonumber\\
 0.0025 \sin{5s})\dots\nonumber
\end{gather}
At $s=0$, the first expansion terms for $r_*$ are
\begin{gather*}
r_*\approx 6 - 0.6111 j^2 - 0.0828 j^4 - 0.211 j^6 -\\
 0.0066 j^8 -0.0023 j^{10}+\dots
\end{gather*}

Substituting the Taylor expansion~(41) in~(39) and~(40) yields Taylor 
expansions for the
binding energy and angular momentum in the marginally stable orbit:
\begin{gather}
1-E_*\approx 0.0572 + 0.0321 j \sin{s}+j^2 (0.0131  
-\nonumber\\
0.0087
\cos{2s}) + j^3 (0.0143 \sin{s} -0.002 \sin{3s}) +\nonumber\\
j^4  (0.0065 - 0.006 \cos{2s} + 0.0003\cos{4s})+\nonumber\\
j^5 (0.0087\sin{s} -0.002\sin{3s})\dots,\nonumber\\
\frac{L_*}{\sin{s}}\approx 3.4641   -0.9428 j \sin{s} - j^2 (0.1123
+\\
0.1443 \sin^2{s})-j^3(0.1178 \sin{s} +0.0175 \sin^3{s}) +\nonumber\\
 j^4 (-0.0138 -0.0878 \sin^2{s} + 0.0138 \sin^4{s}) +\nonumber\\
 j^5 (-0.0327 \sin{s} + 0.0467 \sin^3{s} + 0.016
\sin^5{s})\dots\nonumber
\end{gather}

We supplement expansions~(41) and~(42) with the values of the 
corresponding functions at
$j=1$ numerically constructed by least squares for an arbitrary fixed 
orbital
inclination to the black-hole spin axis:
\begin{gather*}
1-E_*\approx 0.4222 - 0.8314 \cos^2{s} +
0.264\cos^4{s} +\\
0.8605\cos^6{s}-0.656\cos^8{s},
\\
r_*\approx 1+6.2005\cos^4{s} -\\
7.1816\cos^8{s} + 5.1365\cos^{12}{s},\nonumber
\end{gather*}
\begin{gather}
L_*\approx \sin{s}\left(1.1876 +
2.5913\cos^2{s}\right.-\\
\left.0.9948\cos^4{s}+0.494\cos^6{s}\right).\nonumber
\end{gather}
Note that for $s=0$ and $j=1$, the radius of the marginally stable 
orbit is
$1+\sqrt{3}+\sqrt{3+2\sqrt{3}}\approx 5.2745$ (in units of $GM/c^2).$

\section*{Periastron and nodal precession of orbits on the $\quad 
r=\mbox{const}$ surfaces}

Choose the angle~$\theta$ as a parameter on a bounded trajectory. It 
then follows
from~(5)--(7) that
\begin{gather*}
 \frac{d\phi}{d\theta}=\frac{( 2 j E r
+L(\Delta/\sin^2{\theta}-j^2))}{\Delta\sqrt{\Theta}},\\
\Theta\equiv
(1-\frac{\sin^2{s}}{\sin^2{\theta}})\left(\sin^2{\theta} j^2 (1-
E^2)+ \frac{L^2}{\sin^2{s}}\right),
\end{gather*}
\begin{gather} \frac{dt}{d\theta}=\frac{ -2 j r L + 
E(r^2+j^2\cos^2{\theta})\Delta
+2rE(r^2+j^2))}{\Delta\sqrt{\Theta}}.
\end{gather}
Expressions~(39) and~(40) for the energy and angular momentum in 
circular orbits must be
substituted in these formulas for $E$ and $L$.

Integrating the right-hand parts of expressions~(44) over~$\theta$ 
from~$s$ to~$\pi/2$
yields an expression for a quarter of the variation in azimuthal angle 
$\Delta \phi$ and
for a quarter of the period $ T$ in which the particle runs from a 
minimum latitudinal
angle $s$ to its maximum $\pi - s$ and back. The integration result can 
be expressed in
terms of the elliptic functions that contain $E$, $L$, and $s$ as 
parameters [instead of
the constant~$Q$, we inserted the constant~$s$ into formula~(7), which 
has the meaning of
orbital inclination to the black-hole spin axis). Using the notation for
the complete
elliptic integrals of the first, $K(x)$, second, $E(x)$, and third, 
$\Pi(n, x)$, kinds,
we have
\begin{gather*}
 T =4 A
\left( K(k)(j L-j^2 E +(E r^2 +Ej^2 -j
L)\times\right.\\
\left. \frac{(r^2+j^2)}{\Delta})+  (K(k)-E(k)) \frac{E}{
A^2(1-E^2)}\right);
\\
\Delta\phi = 4 A( L\Pi(-\cos^2{s},k) +  (2 r E j - j^2 L) K(k)),
\end{gather*}
\begin{gather}
A\equiv (L^2/\sin^2{s} +j^2(1-E^2))^{-1/2};\\
  k^2\equiv cos^2{s}j^2
(1-E^2) A^2.\nonumber
\end{gather}
In contrast to Johnston and Ruffini
(1974), who first wrote  corresponding formulas in form of elliptic
integrales for any $j$, in formulas~(45), we inserted explicit 
expressions for all
quantities in terms of the turning-point angle $\theta_-\equiv s$ in the 
orbit. Expressions~(39) and~(40) must be substituted 
for $E$ and
$L$, respectively.

For low-eccentricity orbits\footnote{Syer and Clarke (1992) considered 
the possible
existence of stationary disks in the equatorial plane in which the 
particle orbits were
constant-eccentricity ellipses. Clearly, allowance for 
general-relativity effects will
result in the intersection of orbital trajectories in such disks, 
because the orbital
periastron precesses. However, in the Newtonian theory, unstable 
disturbance modes with a
radially variable eccentricity disturbance exist in such disks 
(Lyubarskij \emph{et al.}
1994). Arbitrary accretion-disk disturbances apparently produce spiral 
waves in the disk
structure (Spruit 1987).}($ \delta r = \epsilon\sin{\xi},\quad 
\epsilon << r)$, it
follows from the Hamilton--Jacobi equation~(4) that
\begin{gather}
\frac{d\xi}{d\theta}=\frac{\sqrt{-R^{\prime\prime}/2}}{\sqrt{\Theta}},\quad
R^{\prime\prime}\equiv \frac{d^2 R}{dr^2}=\\ -2\left( 6(1 - E^2)r^2 - 6 r + (1 - E^2)j^2 \sin{s}^2 + \frac{ L^2}{\sin^2{s}}\right).\nonumber
\end{gather}
From Eq.~(46) for the periastron rotation frequency $\nu_r$, we have
\begin{gather}
\omega_r=2\pi\nu_r=\frac{\Delta \xi}{T}= 4 A \frac{\sqrt{R^{\prime\prime}/2}}{T}
K(k).
\end{gather}
Note that the frequencies $\Delta \phi/2\pi T$ and $1/T$ are commonly denoted 
by $\nu_{\phi}$
and $\nu_{\theta}$, respectively, so the sought-for nodal precession 
frequency is
$\nu_{\mathrm{nod}} = \nu_{\phi}-\nu_{\theta}$. The periastron 
precession frequency of
an orbit is $\nu_{\mathrm{per}}=\nu_{\phi}- \nu_r$ (Merloni \emph{et 
al.} 1999).

To clearly present the result, let us derive asymptotic formulas 
from~(45) and~(47) in
the form of Taylor expansions in powers of $j$ by taking into account 
the dependence of
the first integrals $E$, $L$, and $Q$ on $j$ and $s$ given by formulas 
(39), (40), and
(7).

The constants~$E$ and~$\tilde{L}$ enter into the formulas for $\Delta 
\phi$ and $
T$ via the ratios $ E/\tilde{L}$ and $(1-E^2)/L^2,Q/L^2$. Let us write 
out the series
expansions of these ratios with the accuracy that will be required to 
calculate the
nodal precession frequency $(\Delta \phi/2\pi-1)/T$ up to terms of 
the order of $j^3$
inclusive:
\begin{gather*}
\frac{E}{L}=\frac{r-2}{r^{3/2}\sin{s}}+j\frac{3r-4}{r^3}+
\\
j^2\frac{(r+0.5r^2+\sin^2{s}(8+7r-1.5r^2))}{r^{9/2}\sin{s}}+\dots,
\\
\frac{1-E^2}{L^2}=
\frac{r-4}{r^3\sin^2{s}}+j\frac{8(r-2)}{r^{9/2}\sin{s}}+\dots
\end{gather*}
\begin{gather}
\frac{Q}{L^2}=
\cot^2{s}(1+j^2\frac{(r-4)\sin{s}}{2r^3}-{}\\
j^3\frac{4(r-2)\sin{s}}{r^{9/2}}+\dots).\notag
\end{gather}
Using (48), we obtain from expressions~(45) and~(47) for $\Delta \phi$, 
$ T$, and
$\Delta \xi$
\begin{gather}
\frac{\Delta \phi}{2\pi}-1= \frac{2
j}{r^{3/2}}-\frac{j^2}{2r^3}3(r-4)\sin{s}+\nonumber\\
\frac{j^3}{r^{9/2}}(2-1.5r+(18 - 7.5r)\sin^2{s})
+\dots \nonumber \\
\frac{ T}{2\pi}= r^{3/2}+3j\sin{s}+\frac{j^2}{4
r^{3/2}}\times\nonumber \\
(8+3r+(24-9
r)\sin^2{s}) +\dots \nonumber \\
\left(\frac{\Delta \xi}{2\pi}\right)^2= \frac{r-6}{r} + \frac{12
j}{r^{5/2}}(r-2)\sin{s} \frac{j^2}{r^4}\times{}\\
\left(\frac32
r^2+15r-12-\sin^2{s}(r-4)(\frac{15}{2}r - 21)\right)+\dots\nonumber
\end{gather}
In the limit $s \rightarrow 0$, when the spin axis becomes parallel to 
the tangential
plane to the accretion-disk surface, the following formulas hold:
\begin{gather}
\frac{\Delta \phi}{2\pi}-1=\frac{2
j}{r^{3/2}}+\frac{j^3}{r^{9/2}}(2-1.5r)+\nonumber\\
\frac{3 j^{5}}{r^{15/2}}\left(\frac{3}{2} -\frac{7}{4}r
+\frac{19}{32}r^2\right)
+\dots \nonumber \\
\frac{
T}{2\pi}= r^{3/2} +\frac{j^2}{4 r^{3/2}}(8+3r)+{}\nonumber \\
{}+\frac{3 j^4}{64 r^{9/2}} (7 r^2-48 r +80) +\dots \nonumber \\
\left(\frac{\Delta \xi}{2\pi}\right)^2= \frac{r-6}{r} + \frac{3
j^2}{2 r^4}\left( r^2+10r -8)\right.+\\
\left.\frac{3 j^4}{32 r^7}(-352+576 r-286 r^2+9
r^3)\right)+\dots\nonumber
\end{gather}
From formulas~(49) for the Keplerian frequency $ \nu_{\phi}$, the nodal 
precession
frequency $\nu_{\mathrm{nod}}$, and the periastron rotation frequency 
$\nu_{r}$ of a
low-eccentricity orbit around the black-hole spin direction, we derive 
the compact
formulas
\begin{gather*}
\omega_{\phi}=2\pi\nu_{\phi}=\frac{\Delta
\phi}{T}=\frac{1}{r^{3/2}}\left(1+\frac{(2-3\sin{s})
j}{r^{3/2}}\right.
+\\
\left.\frac{((12+9 r) \sin^2{s}-8 -9r)j^2}{4 r^6}+\dots\right);
\\
\omega_{nod}=2\pi\nu_{nod} = (\Delta \phi-2\pi)/ T\approx
\frac{2j}{r^3}-j^2\frac{1.5\sin{s}}{r^{7/2}}+\\
j^3\frac{-2-3r+(6+1.5r)\sin^2{s}}{r^6}+\dots
\end{gather*}
\begin{gather}
(\omega_{r})^2=(2\pi\nu_r)^2 =\frac{r-6}{r^4}+ \frac{6 j}{r^{11/2}}\sin{s}(r+2) 
+\\
 \frac{j^2}{r^7} (12 + 20 r - 3\sin^2{s}(1+r)(10
+r))+\dots\nonumber
\end{gather}
The first term in the formula for nodal precession was found by Lense
and Thirring
(1918) (see also Wilkins 1972). In a coordinate system rotating with the 
frequency
$\nu_{\mathrm{nod}}$, the Keplerian orbit is stationary.

In the limit $ s \rightarrow 0$, when the black-hole spin axis becomes 
tangential to the
disk surface, we can derive the following expansions for the nodal 
precession frequency,
the periastron rotation frequency, and the Keplerian latitudinal
frequency from
formulas~(45) and~(47) using~(50)\footnote{Caution must be exercised 
when passing to the
limit $ s\rightarrow 0$; for example, when $s\rightarrow 0$,
$L/(\sqrt{\Theta}\sin^2{\theta})\rightarrow\delta(\sin{\theta})$, where 
$\delta(x)$
denotes the Dirac $\delta$ function.}
\begin{gather*}
2\pi \nu_{\mathrm{nod}}  =
(\Delta \phi-2\pi)/ T\approx \frac{2j}{r^3}\left(1 -j^2\frac{3
r+2}{2 r^3}\right.+\\
\left.j^4\frac{8+54 r+27 r^2}{16 r^6}+\dots\right)
\\
(2\pi\nu_r)^2 = \frac{r-6}{r^4}+{}\notag\\
\frac{4 j^2}{r^7}-\frac{3 j^4}{r^{10}}(r^3
+128 r^2 +172 r +32)+\dots
\end{gather*}
\begin{gather}
2\pi\nu_{\theta}= \frac{1}{r^{3/2}}-\frac{j^2}{4 r^{9/2}}(3r +
8)+\\
\frac{j^4}{64 r^{15/2}}(16 + 336 r +15 r^2)+\dots\nonumber
\end{gather}
To compare formulas~(51) and~(52) with observations, we should eliminate 
the radius from
them and express the frequencies $\nu_{\mathrm{nod}}$ and $\nu_{r}$ as 
functions of the
Keplerian frequency $\nu_{\phi}$. The functions 
$\nu_{\mathrm{nod}}(\nu_{\phi})$ and
$,\nu_r(\nu_{\phi})$ may be said to be given by~(51) and~(52) in 
parametric form. An
increase in $\nu_2$ and a decrease in the frequency difference between 
the two peaks
$\nu_2-\nu_1$ during observations implies the transition of a radiating 
clump to an
orbit closer to the black hole and the approach of its orbit to the 
marginally stable
orbit. When observations are accumulated, formulas~(51) and~(52) make it 
possible to
determine the orbital inclination to the spin axis of a slowly rotating 
black hole or
neutron star. When analyzing the energy release in the boundary layer 
and in an extended
disk for a rapidly rotating neutron star, we should took into account 
the appearance of
an intrinsic quadrupole moment that exceeds the Kerr one (see 
Sibgatullin and Sunyaev
1998, 2000a, 2000b). We emphasize that formulas~(51) and~(52) are not 
related to the
weak-field approximation and are valid up to the marginally stable 
orbit.

If the disk almost lies in the equatorial plane of a black hole 
($s=\pi/2$), then
formulas~(51) give Taylor expansions of the formulas by Okazaki \emph{et 
al.} (1987) and
Kato (1990) (see also Merloni \emph{et al.} 1999):
\begin{gather*}
 \omega_{\phi}=\frac{1}{r^{3/2} + j},
\\
\omega_{nod}=\frac{1-\sqrt{1-4 j/r^{3/2} + 3 j^2/r^2}}{r^{3/2} + j},
\end{gather*}
\begin{gather}
\omega_r^2 =\frac{1-6/r + 8 j/r^{3/2} -3 j^2/r^2}{(r^{3/2} + j)^2}.
\end{gather}

\subsection*{The nodal frequency of the marginally stable orbit \\ 
\protect for an arbitrary orbital inclination angle $s$}

The following relations (Sibgatullin and Sunyaev 1998, 2000a) hold for
the marginally
stable orbit in the equatorial plane in a Kerr field, in which the 
reciprocal radius of
the marginally stable orbit acts as a parameter:
\begin{gather*}
 j = \frac{4\sqrt{x}-\sqrt{3-2x}}{3x},\quad
E=\sqrt{\frac{3-2x}{3}},\\
L=\frac{2}{3\sqrt{3}x}(2\sqrt{x}\sqrt{3-2x}+ x ),\quad x\equiv 1/r.
\end{gather*}
Let us substitute these expressions in~(53):
\begin{gather*}
\omega_{\phi}=\frac {3 x^{3/2}}{3 -\sqrt{x}\sqrt{3-2x} +4x},\\
\omega_{nod}=\omega_{\phi}(1+\sqrt{\frac23}(\sqrt{x}-\sqrt{3-2x})),\quad
\nu_{r}=0.\tag{53a}
\end{gather*}
If the parameter $x$ is eliminated from~(53a), then we derive the 
following dependence by
least squares:
\begin{gather*}
\omega_{nod} \approx 0.1872
(\omega_{\phi}-\sqrt{6}/36)
+\\
1.7246(\omega_{\phi}-\sqrt{6}/36)^2+1.2064(\omega_{\phi}-\sqrt{6}/36)^3.\tag
{53b}
\end{gather*}
We emphasize that formula~(53b) accurately describes the limiting 
dependence of the nodal
precession frequency on Keplerian frequency at $s=\pi/2$ in the 
marginally stable orbit
over the entire $\omega_{\phi}$ range from~$1/26$ to $0.5$, which 
corresponds to the range
of Kerr parameter from~$-1$ to~1. To derive dimensional dependences, we 
must make the
following substitutions in all formulas: 
$23.1\omega_{\mathrm{nod}}= \nu_{\mathrm{nod}}
(M/1.4 M_{\odot}),23.1\omega_{\phi}= \nu_{\phi} (M/1.4 
M_{\odot}),\dots$, where 
frequencies $\nu_{\phi}, {}\nu_{\mathrm{nod}}$ will be  given hereafter in kHz.

To determine the nodal precession frequency of the marginally stable 
orbit for an
arbitrary orbital inclination~$s$, we use formulas~(45) in which for 
$E,L, r$, we
substitute their expressions in the marginally stable orbit as functions 
of $j$ and $s$,
(41) and (42). Below, we give the highly accurate approximation 
dependences for the
nodal precession frequency and the Keplerian frequency of the marginally 
stable orbit, in
kHz, on Kerr parameter in the entire $j$ range $(-1\leq j\leq 1)$ that 
we derived by
analyzing our numerical calculations for various inclinations of this 
orbit to the
black-hole spin axis:
\begin{gather*}
 s=0:\quad  \nu_{nod}^*\approx \frac{1.4 M_{\odot}}{M}(0.2139 j +
    0.0448
    j^3+\\
    0.0308 j^5 +
    0.007 j^7), \\
 \nu_{\phi}\approx\frac{1.4 M_{\odot}}{M}( 1.5716 +
    0.1933 j + 0.1539 j^2 +\\    
    0.0712j^3 + 0.0852j^4 +
    0.0533 j^5); \\
 s=10^0:\quad  \nu_{nod}^*\approx \frac{1.4 M_{\odot}}{M}(0.2139 j
 +0.0431 j^2 +\\
     0.0674  j^3 + 0.054 j^4 +
 0.0004  j^5 - 0.0414 j^6+\\
       0.038  j^7 + 0.0432 j^8), \\
 \nu_{\phi}\approx\frac{1.4 M_{\odot}}{M}( 1.5716 + 0.3875 j +
    0.2206 j^2  +\\
    0.0675 j^3 +
    0.1681j^4 + 0.1289 j^5); \\
 s= 30^0: \quad \nu_{nod}^*\approx\frac{1.4 M_{\odot}}{M}(0.2139 j
  +\frac{0.1394 j^2}{1 - 0.82 j}), \\
 s= 45^0:\quad \nu_{nod}^*\approx\frac{1.4 M_{\odot}}{M}(0.2139
    j +\frac{0.1923 j^2}{1 - 0.87 j}),
    \\
  s=60^0: \quad \nu_{nod}^*\approx \frac{1.4
    M_{\odot}}{M}(0.2139 j +\frac{0.2499 j ^2}{1 - 0.92 j}), \\
  s=  80^0: \quad \nu_{nod}^*\approx \frac{1.4
    M_{\odot}}{M}(0.2139 j +\frac{0.2535 j ^2}{1 - 0.976 j}).\tag{54}
\end{gather*}
Note that, in contrast to the preceding formulas, the last two formulas 
in~(54)
approximate the numerical data in the range $-1 \leq j \leq 0.99.$


The corresponding limiting dependence of the nodal precession frequency 
on Keplerian
frequency for $s=\pi/6, \pi/4, \pi/3, 4\pi/9$ can be derived from 
numerical calculations
by least squares (here, 
$\tilde{\omega_{\phi}}\equiv\omega_{\phi}-\sqrt{6}/36$):
\begin{gather*}
 s= 30^0: \quad \omega_{nod} \approx 0.3189{}
\tilde{\omega_{\phi}}+ 0.5205(\tilde{\omega_{\phi}})^2  +\\
    4.37
   (\tilde{\omega_{\phi}})^3,\quad 0.049 \leq\omega_{\phi}\leq 0.171;
\\
 s=45^0:\quad  \omega_{nod} \approx 0.2416{} \tilde{\omega_{\phi}}+
1.5953(\tilde{\omega_{\phi}})^2  -\\
    0.2723(\tilde{\omega_{\phi}})^3,\quad 0.044 \leq\omega_{\phi}\leq 0.257;
\\
 s=60^0:\quad\omega_{nod} \approx 0.2031{} \tilde{\omega_{\phi}}+
1.7744(\tilde{\omega_{\phi}})^2  +\\
    0.5149{} (\tilde{\omega_{\phi}})^3,\quad 0.041 \leq\omega_{\phi}\leq 0.414;
\\
 s=80^0:\quad\omega_{nod} \approx 0.1846{} \tilde{\omega_{\phi}} +  1.7929
(\tilde{\omega_{\phi}})^2  +\\
    1.0862{}
    (\tilde{\omega_{\phi}})^3,\quad 0.039 \leq\omega_{\phi}\leq =
0.485.\tag{54a}
\end{gather*}
In formulas~(54a), we give the ranges of Keplerian frequencies that 
correspond to the
range of Kerr parameter from~$-1$ to $+1$. Consider an example. Let the 
black-hole mass
be $2.2 M_{\odot}$ and the Keplerian rotation frequency be $1.2$~kHz.
Since the rotation
frequency in the marginally stable Keplerian orbit of a nonrotating black hole 
with $M=2.2
M_{\odot}$ is $ 23.1\times 1.4/2.2\times\sqrt{6}/36\approx 1$~kHz,
$\tilde{\omega_{\phi}}=0.2\times 2.2/(1.4\times 23.1)\approx 0.0136$. 
According to~(54a), the
nodal precession frequency of the marginally stable orbit is:
$ \nu_{\mathrm{nod}}\approx 42$~Hz for $s=\pi/2$; 
$ \nu_{\mathrm{nod}}\approx
52.6$~Hz for $s=\pi/4$; ${\bf \nu}_{\mathrm{nod}}\approx 65.3$~Hz for 
$s=\pi/6$;
$ \nu_{\mathrm{nod}}\approx 93.6$~Hz for $s=\pi/18$; and 
$ \nu_{\mathrm{nod}}\approx
123$~Hz for $s\rightarrow 0$. We see that the nodal precession frequency 
significantly
depends on the orbital inclination and changes by a factor of~3 as the 
inclination
changes! If the nodal precession frequency is identified with the 
horizontal-branch
oscillation frequency $\nu_{\mathrm{HBO}}$, then it becomes possible 
to determine the
inclination of the marginally stable orbit from the known black-hole 
mass, the Keplerian
frequency, and the nodal precession frequency. At $j<<1$, the nodal precession frequency of 
the marginally
stable orbit is related to the Keplerian frequency by
\begin{gather*}
\omega_{nod}\approx\frac{2}{2+9\sin{s}}\tilde{\omega_{\phi}}.\tag{54b}
\end{gather*}

At a fixed inclination~$s$, the precession frequency~$\nu$ at the 
marginally stable
orbit monotonically increases with~$j$ and reaches a maximum at $j=1$. 
For the mass
$M=1.4 M_{\odot}$ and $s=0$, it is 296~Hz.

The larger the inclination~$s$ at fixed~$j$, the larger the precession 
frequency
$\nu_{\mathrm{nod}}$. As an illustration, let us write out the 
approximation formula for
$\nu_{\mathrm{nod}}$ at $j=0.998$ [Thorne (1974)]\footnote{Studying the evolution
of a black hole under the effect of disk accretion in the equatorial 
plane, Thorne
adduced arguments that the Kerr parameter cannot be larger than $0.998$ 
in this case
when the radiation from inner disk regions is taken into account. Since 
the black hole
mostly captures photons with a negative angular momentum (opposite to 
the black-hole
rotation), the black-hole spinup by disk accretion is counteracted by 
its spindown via
the predominant capture of the photons propagating against the 
rotation.}:
\begin{gather*}
\nu_{nod}\approx 0.29 + 0.7783 \sin{s}+\frac{17.1094 \sin^6{s}}{1+
1.35\sin^4{s}}\mbox{in kHz}.\tag{55}
\end{gather*}

\section*{Conclusions}

According to the ideas explicitly formulated by Bardeen and Petterson 
(1975) and Bardeen
(1977), a geometrically tilted accretion disk is most commonly modeled 
as a set of rings
with the center at the coordinate origin that smoothly turn with 
changing radius under
the effect of viscous torques and gravitomagnetic forces. This approach 
underlies the
existing theories of tilted accretion disks [Papaloizou and Pringle 
(1983) and Markovic
and Lamb (1998, 2000) for small disk deviations from the equatorial 
plane; Pringle
(1992, 1997) for finite disk deviations from the equatorial plane), 
which are based on a
vector equation for the conservation of angular momentum. Some ideas 
developed by
Shakura and Sunyaev (1973, 1976) for flat disks can be used to study the 
physics of
tilted accretion disks. Van~Kerkwijk \emph{et al.} (1998) explained the 
puzzling
spindown and spinup of some X-ray pulsars by the fact that the 
accretion-disk tilt in the
inner regions could become larger than 90 degrees!

The effects considered above take place for the inner nonstationary part
of the disk,
where viscous torques may be disregarded. Matter is not accumulated in 
the marginally
stable orbit because of its instability. In contrast to the statement by 
Stella \emph{et
al.} (1999) that the orbital inclination affects weakly the nodal 
precession frequency,
our detailed analysis shows that this effect is significant. There are 
probably
preferential inclinations of the marginally stable orbits, and there is 
no need to
introduce even harmonics of the nodal frequency to reconcile 
observations with theory.

\section*{Acknowledgments}

I wish to thank Prof. Sunyaev, who drew my attention to the problem 
considered and for
fruitful discussions.


Translated by V. Astakhov
\end{document}